\documentclass[conference,final]{IEEEtran}
\usepackage{geometry}
\usepackage{color}
\usepackage{amsmath}
\usepackage{amsthm}
\usepackage{amssymb}
\usepackage{graphicx}
\usepackage{mathtools}

\usepackage{setspace}
\usepackage{url}

\usepackage{cite}
\usepackage{epsfig}
\usepackage{anysize}
\usepackage{algorithm}
\usepackage{algorithmic}
\usepackage{comment}
\usepackage{setspace} 
\DeclareMathOperator*{\argmin}{\arg\min}
\DeclareMathOperator*{\argmax}{\arg\max}

\makeatletter

\newcommand{\lyxmathsym}[1]{\ifmmode\begingroup\def\b@ld{bold}
  \text{\ifx\math@version\b@ld\bfseries\fi#1}\endgroup\else#1\fi}

\theoremstyle{plain}

\theoremstyle{plain}
\newtheorem{prop}{\protect\propositionname}
\theoremstyle{plain}

\theoremstyle{plain}

\theoremstyle{remark}
\newtheorem{rem}[]{\protect\remarkname}

\marginsize{0.6in}{0.6in}{0.4in}{0.4in}

\makeatother

\providecommand{\remarkname}{Remark}
\providecommand{\lemmaname}{Lemma}
\providecommand{\corollaryname}{Corollary}
\providecommand{\propositionname}{Proposition}
\providecommand{\theoremname}{Theorem}

\ifCLASSOPTIONcompsoc
\usepackage[caption=false,font=normalsize,labelfont=sf,textfont=sf,labelformat=simple]{subfig}
\else
\usepackage[caption=false,font=footnotesize,labelformat=simple]{subfig}
\fi

\begin{document}

\title{Cache-Aided Non-Orthogonal Multiple Access}
\vspace{-1cm}

\author{Lin Xiang, Derrick Wing Kwan Ng, Xiaohu Ge, \\
Zhiguo Ding, Vincent W.S. Wong, and Robert Schober}

\maketitle
\begin{abstract}
In this paper, we propose a novel joint caching and non-orthogonal multiple access (NOMA) scheme to facilitate advanced downlink transmission for next generation cellular networks. In addition to reaping the conventional advantages of caching and NOMA transmission, the proposed cache-aided NOMA scheme also exploits cached data for interference cancellation which is not possible with separate caching and NOMA transmission designs. Furthermore, as caching can help to reduce the residual interference power, several decoding orders are feasible at the receivers, and these decoding orders can be flexibly selected for performance optimization. We characterize the achievable rate region of cache-aided NOMA and investigate its benefits for minimizing the time required to complete video file delivery. Our simulation results reveal that, compared to several baseline schemes, the proposed cache-aided NOMA scheme significantly expands the achievable rate region for downlink transmission, which translates into substantially reduced file delivery times.
\end{abstract}

\section{Introduction}
Wireless caching is a content-centric networking solution to meet the large downlink capacity demands introduced by video streaming in fifth generation (5G) cellular networks \cite{Wang14:Cache,Schober5G,Ge16MCOM:UDN,Xiang17TVT:CLCaching,Xiang16TWC:CoMP,Xiang18TWC:Untrusted}. Recently, caching at streaming user equipments (UEs), e.g. smartphones and tablets, has been advocated \cite{Niesen14IT:CodedCaching,Maddah:ToN15:decentralized} to enhance the streaming quality of experience (QoE) while reducing (i.e., \emph{offloading}) over-the-air traffic. This poses significant challenges for the design of cache placement and delivery as the aggregate cache capacity is distributed across non-cooperating devices with small individual cache memory sizes. Besides, the actual requests of the UEs are difficult to predict during cache placement due to the users' mobility and the random nature of the users' requests.

In the literature, \emph{coded caching} has been proposed as an effective solution for caching at UEs \cite{Niesen14IT:CodedCaching,Maddah:ToN15:decentralized}. By exploiting the cached data as side information, a \emph{coded multicast} format is created for simultaneous error-free video delivery to multiple users, which leads to a multiplicative performance gain that scales with the aggregate cache memory size of the UEs. However, coded caching has an exponential-time computational complexity. Moreover, the caching concepts proposed in \cite{Niesen14IT:CodedCaching,Maddah:ToN15:decentralized} are mainly applicable to noiseless and error-free communication links, e.g. in wireline networks. For wireless networks impaired by fading and noise, however, the performance of coded multicast is limited by the weakest user with the poorest channel condition within the multicast group.

On the other hand, non-orthogonal multiple access (NOMA) is an efficient approach for wireless multiuser transmission that alleviates the adverse effects of fading \cite{SaitoVTCs13:NOMA,Ding17:JSAC:Survey}. Different from multicast and coded multicast, NOMA pairs multiple simultaneous downlink transmissions on the same time-frequency resource via power domain or code domain multiplexing \cite{3GPP:TR36859}. Strong users with favorable channel conditions can cancel the  interference  caused by weak users with poor channel conditions that are paired on the same time-frequency resource, and hence, achieve a high data rate at low transmit powers. Therefore, high transmit powers can be allocated to weak users to achieve communication fairness \cite{Ding16UserP:TVT}. NOMA has also been extended to multicarrier and multi-antenna systems; see \cite{Sun16FD-MC-NOMA:TCOM,Ding16MIMO-NOMA:TWC} and references therein.

So far, wireless caching and NOMA were either investigated separately or combined in a straightforward manner \cite{Ding17Noma:Caching}. For the latter case, NOMA was shown to improve the performance of both caching and delivery in \cite{Ding17Noma:Caching}. In this paper, however, the joint design of caching and NOMA is advocated to maximize the performance gains introduced by caching at UEs. We show that the joint design of caching and NOMA can significantly outperform the straightforward combination of caching and NOMA. To this end, we consider a simple distributed caching scheme for video file delivery. By splitting the video files into several subfiles, superposition transmission of the requested uncached subfiles is enabled during delivery. If the cached content is \emph{hit}, i.e., requested by the caching UE, the proposed cache-aided NOMA scheme enables traditional offloading of the video traffic. Otherwise, the \emph{missed} cached data, which is not requested by the caching UE, is still exploitable as side information to facilitate (partial) interference cancellation for NOMA. The resulting cache-enabled interference cancellation (CIC) can neither be exploited by separate caching and NOMA designs nor by the scheme in \cite{Ding17Noma:Caching}.{

{With CIC, cached data is useful during file delivery even if the users' requests cannot be accurately predicted \emph{a priori}.}} 
Moreover, 
joint CIC and successive interference cancellation (SIC) improves the interference mitigation capability at the UEs and increases the {{number of possible decoding orders }}
 compared to conventional NOMA. 
Furthermore, the performance of both strong and weak users can benefit from CIC. However, adaptive adjustment of the decoding order according to the cache and channel statuses is critical for reaping the benefits of cache-aided NOMA. Hence, we investigate the joint decoding order selection and power and rate allocation optimization problem for minimization of the file delivery time for \emph{fast} video delivery. The main contributions of this paper are as follows:
\begin{itemize}
\item We propose a novel cache-aided NOMA delivery scheme for spectrally efficient downlink transmission. Thereby, cached data is exploited for cancellation of NOMA interference. We characterize the achievable rate region of the proposed scheme. 

\item We jointly optimize the NOMA decoding order and the rate and power allocations for minimization of the delivery time. As the formulated optimization problem is nonconvex, we propose an iterative method to solve it optimally via solving a sequence of convex problems.

\item We show by simulation that the proposed scheme leads to a considerably larger achievable rate region and a significantly reduced delivery time compared to several baseline schemes, including the straightforward combination of caching and NOMA. 
\end{itemize}

\emph{Notations:} $\mathbb{C}$ and $\mathbb{R}_{+}$ denote the sets of complex and nonnegative real numbers, respectively. 
$\mathbb{E}(\cdot)$ is the expectation operator. $\mathcal{CN}\left({\mu},{\sigma}^2\right)$ represents the complex Gaussian distribution with mean  ${\mu}$ and variance ${\sigma}^2$.  $\mathbf{1}\left[\cdot\right]$ denotes an indicator function which is $1$ when the event is true and $0$ otherwise. {{For decoding the received signals, the notation $i\overset{(n)}{\to}x_{f}$ means that $x_{f}$ is the $n$th decoded signal at UE $i$. Similarly, $i\overset{(n)}{\to}(x_{f},\,x_{f'})$ means that  signals $x_{f}$ and $x_{f'}$ are jointly decoded in the $n$th decoding step.}} Finally, $C( \Gamma )\triangleq \log_{2}\left(1+\Gamma\right)$ denotes the capacity function of an additive white Gaussian noise (AWGN) channel, where $\Gamma$ is the signal-to-interference-plus-noise ratio (SINR).

\section{\label{sec2}System Model}

We consider cellular video streaming from a base station (BS) to two UEs indexed by $i$ and $j$, respectively. The BS and the UEs have a single antenna, respectively. UEs $i$ and $j$ request files $W_{A}$ and $W_{B}$ of sizes $V_{A}$ and $V_{B}$ bits, respectively, where $W_{A}\neq W_{B}$. The respective requests are denoted as $(i,A)$ and $(j,B)$. Each UE is equipped with a cache of size $C_{k}$ bits. Thereby, UE $k\in\left\{ i,j\right\} $ can place portions of file $f\in\left\{ A,B\right\} $ into its cache prior to the time of request, e.g. during the early mornings when  cellular traffic is low. As the cache placement is completed before the users' requests are known, the users may cache files which they later do not request. We assume that UE $k$, $k\in \{i, j\}$, has cached $c_{kf}\in[0,1]$ portion of file $W_{f}$, $f \in \left\{ A,B\right\}$. 

\subsection{Cache Status and File Splitting}
Let us define the minimum and maximum portions of content cached for file $W_f$ by 
\begin{equation}
\underline{c}_{f}\triangleq\min_{k\in\left\{ i,j\right\} }c_{kf} \quad \textrm{and}\quad\overline{c}_{f}\triangleq\max_{k\in\left\{ i,j\right\} }c_{kf},
\label{eq1}
\end{equation}
which correspond to the cache status at user 
\begin{equation}
\underline{k}_{f} \!\triangleq\! \argmin_{k\in\left\{ i,j\right\} }\,c_{kf} \;\,\textrm{and} \;\,\overline{k}_{f}\!\triangleq\! \argmax_{k\in\left\{ i,j\right\} }\,c_{kf},\label{eq2}
\end{equation}
$f \in \left\{ A,B\right\}$, respectively. Based on \eqref{eq1} and \eqref{eq2}, four cache configurations are possible at the time of request: 
\begin{itemize}
\item[] Case \mbox{I}: $i=\overline{k}_{B}$ and $j=\overline{k}_{A}$, i.e., $i=\underline{k}_{A}$ and $j=\underline{k}_{B}$; 

\item[] Case \mbox{II}: $i=\overline{k}_{B}$ and $j=\underline{k}_{A}$, i.e., $i=\overline{k}_{A}$ and $j=\underline{k}_{B}$;

\item[] Case \mbox{III}: $i=\underline{k}_{B}$ and $j=\overline{k}_{A}$, i.e., $i=\underline{k}_{A}$ and $j=\overline{k}_{B}$;

\item[] Case \mbox{IV}: $i=\underline{k}_{B}$ and $j=\underline{k}_{A}$, i.e., $i=\overline{k}_{A}$ and $j=\overline{k}_{B}$.

\end{itemize}
In particular, Case \mbox{I} reflects the scenario where the non-requesting user has cached a larger portion of file $W_f$ than the requesting user, which constitutes an unfavorable cache placement for both users but cannot be avoided in practice as user requests cannot be predicted accurately. In the following, due to the limited space, we only consider Case \mbox{I}. However, the derivations for Case \mbox{I} can be extended to Cases \mbox{II}--\mbox{IV} in a relatively straightforward manner.

\begin{figure}[t]
\centering
\includegraphics[width=2.7in]{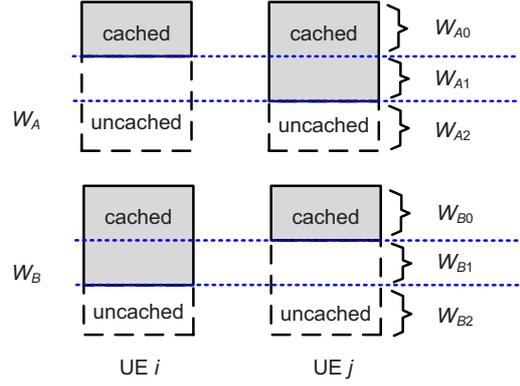}
\vspace{-.2cm}
\caption{\label{fig1a} Illustration of file splitting for cache-aided NOMA, assuming the cache configuration in Case \mbox{I}.} 
\vspace{-.2cm}
\end{figure}

Let $Z_{k}\triangleq(Z_{k,A},Z_{k,B})$, $k \in \{i,j\}$, denote the cache status of UE $k$, where $Z_{k,f}$, $f\in\left\{ A,B\right\}$, is the cached content of file $W_{f}$. We assume that the video data of file $W_{f}$ is sequentially organized. Moreover, based on different user requests and cache configurations, $W_{f}$ is split into three subfiles $(W_{f0},W_{f1},W_{f2})$ for adaptive file delivery. As illustrated in Fig. \ref{fig1a}, $W_{f0}$ and $W_{f2}$ of size $\underline{c}_{f}V_{f}$ and $(1-\overline{c}_{f})V_{f}$ bits are the video chunks which are cached and uncached at both UEs, respectively, whereas subfile $W_{f1}$ of size $(\overline{c}_{f}-c_{kf})V_{f}$ bits is only cached at UE $\overline{k}_{f}$. Hence, we have $Z_{\underline{k}_{f},f}=(W_{f0})$ and $Z_{\overline{k}_{f},f}=(W_{f0},W_{f1})$, $f\in \{A, B\}$. As the cached data $Z_{\underline{k}_{f},f}$ is the prefix of $Z_{\overline{k}_{f},f}$, the considered caching scheme is referred to as \emph{prefix caching} in \cite{Choi:TWC16:Prefix}. 

\subsection{NOMA Transmission}

For video delivery, we assume a frequency flat quasi-static fading channel, where the channel coherence time exceeds the time needed for completion of file delivery. The received signal at UE $k$ is given by 
\begin{equation}
y_{k}=h_{k}x+z_{k},
\end{equation}
where $h_{k}\in\mathbb{C}$ denotes the channel gain between the BS and UE $k$, and is constant during the transmission of file $W_f$. $x$ is the transmit signal and $z_{k}\sim\mathcal{CN}(0,\sigma_{k}^{2})$ is the AWGN at UE $k$.

The BS is assumed to know the cache statuses $Z_{i}$ and $Z_j$ during video delivery. Hence, the BS only transmits the uncached subfiles requested by the UEs. Thereby, for file $W_{f}$, subfiles $W_{f1},\,W_{f2}$, $f\in \{A, B\}$, are encoded employing four independent codebooks at the BS and the corresponding codewords are superposed before being broadcasted over the channel according to the NOMA principle. The resulting BS transmit signal is given by
\begin{equation}
x = \sqrt{p_{i,1}}x_{A1}+\sqrt{p_{i,2}}x_{A2} + \sqrt{p_{j,1}}x_{B1}+\sqrt{p_{j,2}}x_{B2}, 
\label{eq:txsig}
\end{equation}
where $x_{fs}$, $f\in\left\{ A,B\right\}$, $s\in \{1, 2\}$, is the codeword corresponding to subfile $W_{fs}$, and $\mathbb{E} \big[\left|x_{fs}\right|^{2} \big]=1$. Furthermore, $p_{k,s}\ge0$, $k \in \{i,j\}$, $s\in\left\{ 1,2\right\}$, denotes the transmit power of $x_{fs}$.

As the channel is static, we consider time-invariant power allocation, i.e., the powers, $p_{k,s}$, are fixed during file delivery. The total transmit power at the BS is constrained to $P$, i.e., 
\begin{equation}
\textrm{C1:}\;\sum\nolimits _{k\in\left\{ i,j\right\} }\sum\nolimits _{s\in\left\{ 1,2\right\} }p_{k,s}\le P.\label{eq:C1}
\end{equation}
We define $\mathbf{p} \!\triangleq\! (p_{i,1},p_{i,2},p_{j,1},p_{j,2})$ and $\mathcal{P} \!\triangleq\! \left\{ \mathbf{p}\in \mathbb{R}_{+}^{4} \mid \textrm{C1}  \right\}$ as the power allocation vector and the corresponding feasible set, respectively. 

\subsection{Joint CIC and SIC Decoding}

\begin{figure}[t]
\vspace{-.2cm}
\centering
\includegraphics[width=3.2in]{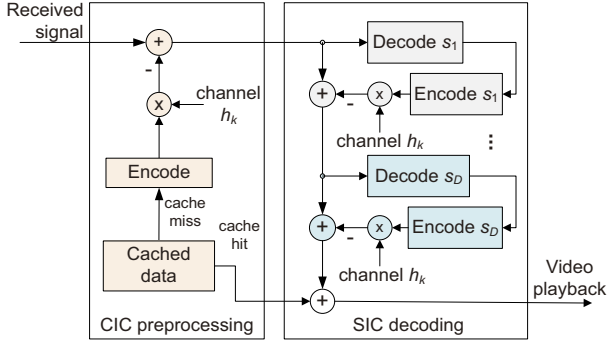}
\vspace{-.2cm}
\caption{\label{fig2a} Joint CIC and SIC decoding at receiver $k\in \{i,j\} $ for cache-aided NOMA.  {{$s_1, \ldots, s_D$, $D \le 4$, represent the residual signals $x_{fs}$, $f\in \{A,B\}$, $s \in \{1,2\}$, which are not canceled by CIC but decoded successively by employing SIC.}}}
\vspace{-.2cm}
\end{figure}

The proposed cache-aided NOMA scheme enables CIC at the receiver, which is not possible for conventional NOMA. The joint CIC and SIC receiver performs CIC preprocessing of the received signal before SIC decoding as illustrated in Fig.~\ref{fig2a}. In particular, the interference caused by transmit signal $x_{A1}$ ($x_{B1}$), which is requested by UE $i$ (UE $j$), can be canceled at UE $j$ (UE $i$) by exploiting the cached data $Z_{j,A}$ ($Z_{i,B}$). Hence, the residual received signal after CIC preprocessing is given by
\begin{alignat}{1}
y_{i}^{\mathrm{CIC}} &= h_{i}(\sqrt{p_{i,1}}x_{A1}+\sqrt{p_{i,2}}x_{A2}+\sqrt{p_{j,2}}x_{B2})+z_{i}, \label{eq9} \\
y_{j}^{\mathrm{CIC}}  &=h_{j}(\sqrt{p_{i,2}}x_{A2}+\sqrt{p_{j,1}}x_{B1}+\sqrt{p_{j,2}}x_{B2})+z_{j}.  \label{eq10} 
\end{alignat}

\begin{rem}
For the proposed cache-aided NOMA scheme, $c_{kf}$ portion of file $W_f$ is not transmitted at all, while $\overline{c}_{f}-c_{kf}$ portion of file $W_f$ can be removed from the received signal of the non-requesting UE $k'$ via CIC, where $(k,f)\in\left\{ (i,A),(j,B)\right\}$ and $k'\neq k$. As such, the proposed scheme can exploit $\overline{c}_f$ portion of $W_f$ for performance improvement, even for the unfavorable cache configuration of Case \mbox{I}, whereas a straightforward combination of caching and NOMA can only exploit the cached portion $c_{kf}$  of the requested file \cite{Ding17Noma:Caching}. 
\end{rem}

As CIC reduces the multiuser interference power, multiple decoding orders become possible for SIC processing of $y_{k}^{\mathrm{CIC}}$. For example, there are $4!=24$ possible decoding orders based on \eqref{eq9} and \eqref{eq10}, compared to $2!=2$ for conventional NOMA. This leads to a substantially increased flexibility in decoding the video data based on $y_{k}^{\mathrm{CIC}}$. 

Optimizing the SIC decoding order and the respective transmission rates and power allocations based on the cache status and channel conditions enhances the performance of video delivery. On the other hand, decoding order optimization is a combinatorial problem, which may increase complexity. However, by careful inspection of the SIC decoding conditions, we show in Section~\ref{sec3} that the optimal decoding order is contained in a small subset of all possible decoding orders, and hence the associated complexity is limited. 

\section{\label{sec3}Achievable Rate Region and Delivery Time Minimization}

In this section, we evaluate the achievable rate region of the proposed cache-aided NOMA scheme. Based on the derived results, we then minimize the delivery time during file transfer by optimizing the decoding order and the power and rate allocation. Let $\mathbf{r}\triangleq(r_{i,1},r_{i,2},r_{j,1},r_{j,2})$ be the rate allocation vector, where  $r_{k,s}\ge0$ is the rate for delivering $x_{fs}$ to UE $k$, $(k,f) \in \{(i,A),(j,B)\}$, $s\in\left\{ 1,2\right\}$. We define $\alpha_{k}\triangleq\frac{\sigma_{k}^{2}}{\left|h_{k}\right|^{2}}$, $k \in \{i, j\}$, as the effective noise variance at UE $k$. Without loss of generality, we assume $\alpha_{i}<\alpha_{j}$, i.e., UE $i$ has a larger channel gain than UE $j$. 

\subsection{\label{sub3-1}Derivation of Achievable Rate Region}

According to \eqref{eq9} and \eqref{eq10}, two subfiles, $x_{f1}$ and $x_{f2}$, are delivered to each user and $x_{B1}$ ($x_{A1}$) is canceled at UE $i$ ($j$) by CIC. Moreover, $x_{A2}$ and $x_{B2}$, which are interference signals at one user, are commonly received at both users, whereas $x_{A1}$ and $x_{B1}$ are received only at the requesting users. The interference signals can be decoded and canceled only if the SIC decoding condition is fulfilled, i.e., the received SINR for $x_{A2}$ and $x_{B2}$ at the non-requesting users, UE $j$ and UE $i$, has to exceed that at the requesting users, UE $i$ and UE $j$, respectively. In contrast, signals $x_{A1}$ and $x_{B1}$ can be decoded without such constraint. Depending on which signal is decoded first, three cases can be distinguished: for the first two cases, signals $x_{A1}$ and $x_{B1}$ are decoded first at the requesting users, respectively, whereas, for the third case, the interference signals $x_{A2}$ and $x_{B2}$ are decoded first. For these cases, as \eqref{eq9} and \eqref{eq10} constitute a non-degraded broadcast channel, the corresponding achievable rate regions have to be evaluated for specific power regions individually. 

\subsubsection{\label{enu:IV-1} UE $i\overset{(1)}{\to}x_{A1}$}
 If UE $i$ decodes $x_{A1}$ first, signals $y_{i}^{\mathrm{(1)}}=h_{i}(\sqrt{p_{i,2}}x_{A2}+\sqrt{p_{j,2}}x_{B2})+z_{i}$ and $y_{j}^{\mathrm{(1)}}=h_{j}(\sqrt{p_{i,2}}x_{A2}+\sqrt{p_{j,2}}x_{B2}+\sqrt{p_{j,1}}x_{B1})+z_{j}$ have to be decoded subsequently. The achievable rate region is provided in Proposition \ref{prop1}.

\vspace{-.3cm}
\begin{prop}
\emph{\label{prop1} When UE $i\overset{(1)}{\to}x_{A1}$, the rate region $ \mathcal{R}_{1}\left(\mathcal{P}_{1}\right)\bigcup\mathcal{R}_{2}\left(\mathcal{P}_{2}\right)$ is achievable, where 
\vspace{-.1cm}
\begin{alignat}{1}
&\mathcal{R}_{1}\left(\mathcal{P}_{1}\right) \! \triangleq \!\!\bigcup_{\mathbf{p}\in\mathcal{P}_{1}}\!\! \left\{ \mathbf{r}\left|\begin{array}{l}
r_{i,1}\le C_{i,1}=C\left(\frac{p_{i,1}}{p_{i,2}+p_{j,2}+\alpha_{i}}\right) \\
 r_{i,2}\le C_{i,2}=C\left(\frac{p_{i,2}}{\alpha_{i}}\right) \\
r_{j,s}\le C_{j,s}=C\left(\frac{p_{j,s}}{p_{i,2}+\alpha_{j}}\right), s= 1,2 \\
r_{j,1}+r_{j,2}\le C_{j,1,2}=C\left(\frac{p_{j,1}+p_{j,2}}{p_{i,2}+\alpha_{j}}\right)  
\end{array} \!\!  \right. \right\} \nonumber \\ 
&\mathcal{R}_{2}\left(\mathcal{P}_{2}\right) \! \triangleq \! \! \bigcup_{\mathbf{p}\in\mathcal{P}_{2}} \!\! \left\{ \mathbf{r}\left|\begin{array}{l}
r_{i,1}\le C_{i,1}=C\left(\frac{p_{i,1}}{p_{i,2}+p_{j,2}+\alpha_{i}}\right) \\
 r_{i,2}\le C_{i,2}=C\left(\frac{p_{i,2}}{p_{j,2}+\alpha_{i}}\right) \\
r_{j,1}\le C_{j,1}=C\left(\frac{p_{j,1}}{\alpha_{j}}\right) \\
 r_{j,2}\le C_{j,2}=C\left(\frac{p_{j,2}}{p_{i,2}+p_{j,1}+\alpha_{j}}\right) 
\end{array}\right.\right\} \nonumber 
\end{alignat}
with $\mathcal{P}_{1} = \mathcal{P}$ and $\mathcal{P}_{2}\triangleq\left\{ \mathbf{p}\in\mathcal{P}\mid p_{j,2}-p_{j,1}>\alpha_{j}-\alpha_{i}\right\} $. For $\mathcal{R}_{1}\left(\mathcal{P}_{1}\right)$, the decoding orders for UEs $i$ and $j$ are given as $i\overset{(1)}{\to}x_{A1}\overset{(2)}{\to}x_{B2}\overset{(3)}{\to}x_{A2}$ and $j\overset{(1)}{\to}\left(x_{B1},\,x_{B2}\right)$, respectively.  For $\mathcal{R}_{2}\left(\mathcal{P}_{2}\right)$, the decoding orders are $i\overset{(1)}{\to}x_{A1}\overset{(2)}{\to}x_{A2}$ and $j\overset{(1)}{\to}x_{A2}\overset{(2)}{\to}x_{B2}\overset{(3)}{\to}x_{B1}$, respectively. }
\end{prop} 
\vspace{-.2cm}
\begin{IEEEproof}
Please refer to Appendix \ref{proof1}.
\end{IEEEproof}

\begin{rem}
In Proposition~\ref{prop1}, the interference for decoding $x_{A2}$ is reduced after $x_{A1}$ has been decoded and canceled from $y_{i}^{\mathrm{CIC}}$. Hence, decoding $x_{A1}$ first is desirable when e.g. $W_{A1}$ has a smaller size and/or requires a lower delivery rate than $W_{A2}$. The decoding orders for $\mathcal{R}_{1}\left(\mathcal{P}_{1}\right)$ favor the delivery of $x_{A2}$ to UE $i$ as it experiences no interference after SIC, and thus can attain a high data rate $r_{i,2}$ even for small transmit powers $p_{i,2}$. In contrast, the decoding orders for $\mathcal{R}_{2}\left(\mathcal{P}_{2}\right)$ favor the delivery of $x_{B1}$ to UE $j$. This implies that $\mathcal{R}_{2}\left(\mathcal{P}_{2}\right)$  expands $\mathcal{R}_{1}\left(\mathcal{P}_{1}\right)$ along $r_{j,1}$. 
\end{rem}

\subsubsection{\label{enu:IV-2} UE $j\overset{(1)}{\to}x_{B1}$ excluding UE $i\overset{(1)}{\to}x_{A1}$}

\footnote{The achievable rate region for UE $j\overset{(1)}{\to}x_{B1}$ and UE $i\overset{(1)}{\to}x_{A1}$ is already included in $\mathcal{R}_{1}\left(\mathcal{P}_{1}\right)$, and hence, excluded herein.}Decoding and canceling $x_{B1}$ first improves the SINR of $x_{B2}$ at UE $j$, which is desirable when subfile $W_{B1}$ has a smaller size than $W_{B2}$. The resulting signals after $x_{B1}$ has been canceled are $y_{i}^{\mathrm{(1)}} = h_{i}(\sqrt{p_{i,1}}x_{A1}+\sqrt{p_{i,2}}x_{A2}+\sqrt{p_{j,2}}x_{B2}) + z_{i}$ and $y_{j}^{\mathrm{(1)}} = h_{j}(\sqrt{p_{i,2}}x_{A2}+\sqrt{p_{j,2}}x_{B2}) + z_{j}$. The corresponding achievable rate region is given in Proposition~\ref{prop2}.

\vspace{-.3cm}
\begin{prop}
\emph{\label{prop2} When UE $j\overset{(1)}{\to}x_{B1}$ but UE $i\overset{(1)}{\to}x_{A1}$ is excluded, the achievable rate region is given by $\mathcal{R}_{3}\left(\mathcal{P}_{3}\right)\bigcup\mathcal{R}_{4}\left(\mathcal{P}_{4}\right)$, where 
\vspace{-.1cm}
\begin{alignat}{1}
 & \mathcal{R}_{3}\left(\mathcal{P}_{3}\right) \! \triangleq \! \! \bigcup_{\mathbf{p}\in\mathcal{P}_{3}} \!\! \left\{ \mathbf{r}\left|\begin{array}{l}
r_{i,s}\le C\left(\frac{p_{i,s}}{\alpha_{i}}\right),\,s= 1,2 \\
r_{i,1}+r_{i,2}\le C\left(\frac{p_{i,1}+p_{i,2}}{\alpha_{i}}\right) \\
r_{j,1}\le C\left(\frac{p_{j,1}}{p_{i,2}+p_{j,2}+\alpha_{j}}\right) \\
 r_{j,2}\le C\left(\frac{p_{j,2}}{p_{i,2}+\alpha_{j}}\right) \\
\end{array} \!\! \right.\right\}  \nonumber \\ 
 & \mathcal{R}_{4}\left({P}_{4}\right)\! \triangleq \!\!\bigcup_{\mathbf{p}\in \mathcal{P}_{4}} \!\! \left\{ \mathbf{r}\left|\begin{array}{l}
r_{i,1}\le C\left(\frac{p_{i,1}}{p_{j,2}+\alpha_{i}}\right) \\
r_{i,2}\le C\left(\frac{p_{i,2}}{p_{i,1}+p_{j,2}+\alpha_{i}}\right) \\
r_{j,1}\le C\left(\frac{p_{j,1}}{p_{i,2}+p_{j,2}+\alpha_{j}}\right) \\
r_{j,2}\le C\left(\frac{p_{j,2}}{\alpha_{j}}\right) \\
\end{array} \!\! \right.\right\}  \nonumber  
\end{alignat}
with $\mathcal{P}_{3}\triangleq\left\{ \mathbf{p}\in\mathcal{P}\mid p_{i,1}<\alpha_{j}-\alpha_{i}\right\} $ and $\mathcal{P}_{4} = \mathcal{P}\backslash\mathcal{P}_{3}$. The decoding orders achieving $\mathcal{R}_{3}\left(\mathcal{P}_{3}\right)$ are UE $i\overset{(1)}{\to}x_{B2}\overset{(2)}{\to}(x_{A1},x_{A2})$ and UE $j\overset{(1)}{\to}x_{B1}\overset{(2)}{\to}x_{B2}$.  Moreover, the decoding orders for $\mathcal{R}_{4}\left(\mathcal{P}_{4}\right)$ are UE $i\overset{(1)}{\to}x_{A2}\overset{(2)}{\to}x_{A1}$ and UE $j\overset{(1)}{\to}x_{B1}\overset{(2)}{\to}x_{A2}\overset{(3)}{\to}x_{B2}$.} 
\end{prop}
\vspace{-.2cm}
\begin{IEEEproof}
Please refer to Appendix \ref{proof2}.
\end{IEEEproof}


\subsubsection{\label{enu:IV-3}UE $j\overset{(1)}{\to}(x_{A2},x_{B2})$ and UE $i\overset{(1)}{\to}(x_{A2},x_{B2})$}
Recall that decoding the interference signals first is only possible if the SIC condition is fulfilled. In this case, the achievable rate region is given in Proposition \ref{prop3}.

\vspace{-.3cm}
\begin{prop}
\emph{\label{prop3} When UE $j\overset{(1)}{\to}(x_{A2},x_{B2})$ and UE $i\overset{(1)}{\to}(x_{A2},x_{B2})$, the achievable rate region is given by $\mathcal{R}_{5}\left(\mathcal{P}_{5}\right)\bigcup\mathcal{R}_{6}\left(\mathcal{P}_{6}\right)\bigcup\mathcal{R}_{7}\left(\mathcal{P}_{7}\right)$, where
\vspace{-.1cm}
\begin{alignat}{1}
 & \mathcal{R}_{5}\left(\mathcal{P}_{5}\right) \!\triangleq\! \! \bigcup_{\mathbf{p}\in\mathcal{P}_{5}} \! \left\{ \mathbf{r}\left|\begin{array}{l}
r_{i,s}\le C\left(\frac{p_{i,s}}{\alpha_{i}}\right),\,s\in\left\{ 1,2\right\}  \\
 r_{i,1}+r_{i,2}\le C\left(\frac{p_{i,1}+p_{i,2}}{\alpha_{i}}\right)  \\
r_{j,1}\le C\left(\frac{p_{j,1}}{p_{i,2}+\alpha_{j}}\right) \\
 r_{j,2}\le C\left(\frac{p_{j,2}}{p_{j,1}+p_{i,2}+\alpha_{j}}\right) 
\end{array} \!\! \right.\right\} \nonumber \\ 
 & \mathcal{R}_{6}\left(\mathcal{P}_{6}\right)\! \triangleq \!\! \bigcup_{\mathbf{p}\in \mathcal{P}_{6}} \! \left\{ \mathbf{r}\left|\begin{array}{l}
r_{i,1}\le C\left(\frac{p_{i,1}}{\alpha_{i}+p_{j,2} \Delta }\right)  \\
r_{i,2}\le C\left(\frac{p_{i,2}}{p_{i,1}+p_{j,2}+\alpha_{i}}\right)  \\
r_{j,1}\le C\left(\frac{p_{j,1}}{\alpha_{j}}\right)  \\
r_{j,2}\le C\left(\frac{p_{j,2}}{p_{i,2}+p_{j,1}+\alpha_{j}}\right)  
\end{array} \!\! \right.\right\} \nonumber \\ 
 & \mathcal{R}_{7}\left(\mathcal{P}_{7}\right)\!\triangleq\! \bigcup_{\mathbf{p}\in \mathcal{P}_{7}} \!\! \left\{ \mathbf{r}\left|\begin{array}{l}
r_{i,1}\le C\left(\frac{p_{i,1}}{p_{j,2}+\alpha_{i}}\right)  \\
r_{i,2}\le C\left(\frac{p_{i,2}}{p_{i,1}+p_{j,2}+\alpha_{i}}\right)  \\
r_{j,s}\le C\left(\frac{p_{j,s}}{\alpha_{j}}\right),\,s= 1,2  \\
 r_{j,1}+r_{j,2}\le C\left(\frac{p_{j,1}+p_{j,2}}{\alpha_{j}}\right)  
\end{array} \!\!\! \right.\right\} \nonumber 
\end{alignat}
with $\mathcal{P}_{5}\triangleq\left\{ \mathbf{p}\in\mathcal{P}\mid p_{i,1}<p_{j,1}+\alpha_{j}-\alpha_{i}\right\}$, $\mathcal{P}_{6} = \mathcal{P}_{7} = \mathcal{P}\backslash\mathcal{P}_{5}$, and $\Delta \triangleq \mathbf{1}\left[p_{i,2} > p_{i,1} - p_{j,1} -\alpha_{j}+\alpha_{i}\right]$.
The decoding orders achieving $\mathcal{R}_{5}\left(\mathcal{P}_{5}\right)$ are UE $i\overset{(1)}{\to}x_{B2}\overset{(2)}{\to}(x_{A1},x_{A2})$ and UE $j\overset{(1)}{\to}x_{B2}\overset{(2)}{\to}x_{B1}$. Moreover, $\mathcal{R}_{6}\left(\mathcal{P}_{6}\right)$ is achieved by the decoding orders
\[
\textrm{UE}\;i\overset{(1)}{\to}x_{A2}\overset{(2)}{\to}\begin{cases}
x_{A1}, & \textrm{if } \Delta = 1,\\
x_{B2}\overset{(3)}{\to}x_{A1}, & \textrm{otherwise,}
\end{cases}
\]
 and UE $j\overset{(1)}{\to}x_{B2}\overset{(2)}{\to}x_{B1}$. Finally, $\mathcal{R}_{7}\left(\mathcal{P}_{7}\right)$ is achieved by the decoding orders UE $i\overset{(1)}{\to}x_{A2}\overset{(2)}{\to}x_{A1}$ and UE $j\to x_{A2}\overset{(2)}{\to}(x_{B1},x_{B2})$.}
\end{prop}
\vspace{-.2cm}
\begin{IEEEproof}
Please refer to Appendix \ref{proof3}.
\end{IEEEproof}

{{\begin{rem}
Different from conventional NOMA, file splitting in the proposed cache-aided NOMA scheme enables joint decoding opportunities. For example, joint decoding of $x_{B1}$ and $x_{B2}$ at UE $j$ is possible in $\mathcal{R}_{1}\left(\mathcal{P}_{1}\right)$, as the two signals are received by UE $j$ over the same AWGN channel with noise variance $p_{i,2} + \alpha_j$. Therefore, UE $j$ can choose the decoding order for these files without restriction. Similarly, joint decoding of $x_{A1}$ and $x_{A2}$ is possible at UE $i$ in $\mathcal{R}_{3}\left(\mathcal{P}_{3}\right)$ and $\mathcal{R}_{5}\left(\mathcal{P}_{5}\right)$, respectively. We note that employing file splitting in conventional NOMA would not increase the achievable rates at UEs. However, if a portion of file is cached at one of the UEs, the achievable rates of the UEs can be increased by employing file splitting in the proposed cache-aided NOMA as this enables CIC.   
\end{rem} }}

Finally, combining the results in Propositions \ref{prop1}--\ref{prop3}, the overall achievable rate region is $\mathcal{R}  \triangleq \bigcup_{n=1}^{7} \mathcal{R}_{n} (\mathcal{P}_{n})$. Note that $\mathcal{R}_{n} (\mathcal{P}_{n})$ can be written in general form as
\begin{equation}
\mathcal{R}_{n}(\mathcal{P}_{n}) \!\!=\!\! \bigcup_{\mathbf{p} \in \mathcal{P}_{n}} \!\!\! \left\{ \mathbf{r}\left| \!\!
\begin{array}{l}
\textrm{C2: } r_{ks} \! \le \! C_{k,s}(\mathbf{p}), k \! \in \! \left\{\! i,j \! \right\}, s \!\in \! \left\{\! 1,2 \!\right\}   \\
\textrm{C3: } r_{k1} \!+\! r_{k2} \!\le\! C_{k,1,2}(\mathbf{p}), k \!\in \! \left\{\! i,j \!\right\}  
\end{array} \!\!\!\!  \right.\right\} 
\label{eq:rateoverall}
\end{equation}
where $C_{k,s}$ and $C_{k,1,2}$ are the respective capacity bounds for decoding signal $x_{fs}$, $s\in \{1,2\}$, and signals $\left\{ x_{f1},x_{f2}\right\} $ at user $k\in\left\{ i,j\right\}$.

\subsection{\label{sec4}Rate and Power Allocation for Fast Delivery}
 
Let $T$ be the time required to complete the delivery of the requested files. We have
\begin{equation}
T=\max_{k\in\left\{ i,j\right\} ,\,s\in\left\{ 1,2\right\}}\;\frac{\beta_{k,s}}{r_{k,s}},\label{eq:del-time}
\end{equation}
for $\mathbf{r}\in\mathcal{R}$, where $\beta_{k,1}\triangleq(\overline{c}_{f}-c_{kf})V_{f}$ and $\beta_{k,2}\triangleq(1-\overline{c}_{f})V_{f}$ for $(k,f)\in\left\{ (i,A),(j,B)\right\}$ denote the effective volume of data to be delivered to user $k$. To avoid trivial results, we assume throughout this section that $\beta_{k,1}+\beta_{k,2}>0$, $\forall k\in\left\{ i,j\right\} $, i.e., each user requests some video data that is not cached\footnote{Otherwise, $p_{k,1}=p_{k,2}=0$ and $r_{k,1}=r_{k,2}=0$.}. Consequently, the delivery time optimization problem is formulated as 
\vspace{-.2cm}
\begin{alignat}{1}
\textrm{P1:}\; \min_{\mathbf{r}\in\mathcal{R},\;\mathbf{p}\in\mathcal{P},\;T\ge0}\; & T\\
\mathrm{s.t.}\quad\quad\; & \textrm{C4:}\;r_{ks}T\ge\beta_{k,s},\; k\in\left\{ i,j\right\}, s\in\left\{ 1,2\right\}, \nonumber 
\end{alignat}
where C4 ensures completion of file delivery at time $T$.

Problem P1 is generally nonconvex as the capacity functions in C2 and C3 in \eqref{eq:rateoverall} are not jointly convex with respect to $\mathbf{r}$ and $\mathbf{p}$, and C4 is bilinear. However, the optimal solution of Problem P1 can be obtained by solving a sequence of convex problems as will be shown in the following. In particular, {{assume that the optimal solution lies in $\mathcal{R}_{n}$.}} 
For each feasible power allocation, the rate region $\mathcal{R}_{n}$, cf. \eqref{eq:rateoverall}, reduces to a polyhedron. Consequently, the optimal rate allocation, denoted as $\mathbf{r}^* \triangleq (r_{i,1}^*,r_{i,2}^*,r_{j,1}^*,r_{j,2}^*)$, can be obtained as the rate tuple on the dominate face\footnote{For a polyhedron, any point that lies outside the dominant face is dominated elementwise by some point on the dominant face \cite{Tse2005Fundamentals}.}
of $\mathcal{R}_{n}$ \cite{Tse2005Fundamentals}. For example, for $n=1$, we have $r_{i,1}^{*} =C_{i,1}$ and $r_{i,2}^{*} =C_{i,2}$ as the rates of UE $i$ are only constrained by C2. In contrast, as the rates of UE $j$ are constrained by both C2 and C3, we have $r_{j,1}^{*}=C\left(\frac{p_{j,1}^{*}}{p_{j,2}^{*}+p_{i,2}^{*}+\alpha_{j}}\right)$ and $ r_{j,2}^{*}=C\left(\frac{p_{j,2}^{*}}{p_{i,2}^{*}+\alpha_{j}}\right)$ for decoding order $j\overset{(1)}{\to} x_{B1}\overset{(2)}{\to}x_{B2}$ and $r_{j,1}^{*}=C\left(\frac{p_{j,1}^{*}}{p_{i,2}^{*}+\alpha_{j}}\right)$ and $ r_{j,2}^{*}=C\left(\frac{p_{j,2}^{*}}{p_{i,2}^{*}+p_{j,1}^{*}+\alpha_{j}}\right)$ for decoding order $j\overset{(1)}{\to} x_{B2}\overset{(2)}{\to}x_{B1}$, and the optimal power allocation $\mathbf{p}^* =  (p_{i,1}^*,p_{i,2}^*,p_{j,1}^*,p_{j,2}^*) $. In the same manner, the optimal rate allocation for all $n\in \{1,\ldots,7\}$ can be obtained.

Substituting the optimal rate allocations {{and letting $\rho_n = 1/T$,}} P1 can be equivalently reformulated as  $T^* = \min_{n\in\left\{ 1,\ldots,7\right\} }\rho_{n}^{*}$ with
\vspace{-.2cm}
\begin{alignat}{1}
\rho_{n}^{*}\triangleq\max_{\mathbf{p}\in\mathcal{P}_{n},\,\rho_{n}\ge0}\; & \rho_{n}\label{eq:61}\\
\mathrm{s.t.}\quad\; & {\textrm{C5: }} r_{k,s}^{*}\left(\mathbf{p}\right)\ge\rho_{n}\beta_{k,s}, k \!\in\! \left\{\! i,j \!\right\} , s \!\in\! \left\{\! 1,2 \! \right\} .\nonumber 
\end{alignat}
The optimal value $\rho_{n}^{*}$ can be found iteratively by employing Algorithm~\ref{alg1}. In particular, in each iteration, the feasibility of problem \eqref{eq:61} is checked for a given $\rho_{n}$, cf. line~\ref{alg1:line9}. For given $\rho_{n}$, we have $\rho_{n}^{*}\ge \rho_{n}$ if \eqref{eq:61} is feasible, i.e., $\rho_{n}$ is a lower bound on $\rho_{n}^{*}$, and $\rho_{n}^{*}\le \rho_{n}$ otherwise, i.e., $\rho_{n}$ is an upper bound on $\rho_{n}^{*}$. Hence, a bisection search can be applied to iteratively update the value of $\rho_{n}$ until the gap between the lower and the upper bounds vanishes, whereby $\rho_{n}^{*}$ is obtained. Moreover, efficient convex optimization algorithms can be employed \cite{Boyd2004Convex} in line~\ref{alg1:line9} of Algorithm~\ref{alg1}. This is because although C5 is a linear fractional constraint of the form $\log_{2}\left(1+\frac{\mathbf{a}^{T}\mathbf{p}}{\mathbf{b}^{T}\mathbf{p}+1}\right)\ge c$ for $\mathbf{a}, \mathbf{b}\in \mathbb{R}_+^4$ and $c \in \mathbb{R}_+$, it can be transformed into an equivalent convex constraint of the form $\left(\mathbf{a}-\left(2^{c}-1\right)\mathbf{b}\right)^{T}\mathbf{p}\ge2^{c}-1$ such that an equivalent convex formulation of problem \eqref{eq:61} is obtained. 

\begin{algorithm}[t]
\protect\caption{\textcolor{black}{Bisection search for $\rho_n^*$.} }

\label{alg1} 
\small{
\begin{algorithmic}[1]
\STATE \textbf{initialization}: {Given} $LB$, $UB$, and tolerance $\epsilon$;  

\REPEAT 
\STATE $\rho_n \leftarrow  (LB+UB)/2   $;
\STATE  Solve the feasibility problem of \eqref{eq:61} for $\rho_n $;\label{alg1:line9}
\IF{\eqref{eq:61} is infeasible}{\STATE $UB \leftarrow \rho_n$; } \ELSE{ \STATE $LB \leftarrow \rho_n$; } \ENDIF
\UNTIL{$UB - LB < \epsilon$.} \label{alg1:line17} 
\end{algorithmic} 
}
\end{algorithm}

\section{\label{sec5}Performance Evaluation}

In this section, the performance of the proposed cache-aided NOMA is evaluated by simulation. Consider a cell of radius $R=2$~km, where the BS is deployed at the center of the cell and the strong and the weak users, UE $i$ and UE $j$, are uniformly distributed on discs of radii $R_i = 0.2$~km and $R_j = 0.6$~km, respectively. For modeling the wireless channel, the 3GPP path loss model (``Urban Macro NLOS'' scenario) in \cite{3GPP:TR36814} is adopted. The small-scale fading coefficients are independent and identically distributed (i.i.d.) Rayleigh random variables. The video files have size $V_{A} = V_{B} =$500~MBytes. Moreover, the system has a bandwidth of $5$~MHz. The noise power spectral density is $-$172.6 dBm/Hz. Finally, we set 
the maximal transmit power as $P = 35$ dBm and the cache status as $c_{iA} = 0.2$, $c_{iB} = 0.8$, $c_{jA} = 0.8$, and $c_{jB} = 0.2$.


\subsection{\label{sub3-4}Baseline Schemes}

\subsubsection{Baseline 1 (Cache-aided orthogonal multiple access (OMA))}

As baseline, we consider time-division multiple access (TDMA) for transmitting the uncached portions of the requested files. In particular, $\tau$ and $1-\tau$ fractions of time are allocated for transmission to UE $i$ and UE $j$, respectively, where $\tau \in [0,1]$. Consequently, the capacity region for all possible time allocations is given by $ \mathcal{R}_{\mathrm{OMA}} \!= \!\!\bigcup_{\tau \in [0,1] }  \!\left\{ (r_{i},r_{j}) \left|\!\! \begin{array}{c} r_{i}\le\tau C(\frac{P}{\alpha_{i}}), r_{j}\le (1-\tau) C(\frac{P}{\alpha_{j}})\end{array} \!\!\!\! \right.\right\}$. Note that, with Baseline 1, caching only facilitates conventional offloading of the hit cached data.

\subsubsection{Baseline 2 (Conventional NOMA with and without caching)}

If caching is possible, Baseline 2 is a straightforward combination of caching and NOMA, whereby the requested data hit by the cache is offloaded and only the remaining data is transmitted by applying NOMA. If caching is not possible, Baseline 2 reduces to the conventional NOMA scheme. In both cases, the BS transmits signals $x=\sqrt{p_{i}}x_{A}+\sqrt{p_{j}}x_{B}$ for delivering files $W_A$ and $W_B$, where the power allocations $p_{i}$ and $p_{j}$ satisfy $\mathcal{P}_{\mathrm{NOMA}}\triangleq\left\{ (p_{i},p_{j})\in\mathbb{R}_{+}^{2}\mid p_{i}+p_{j}\le P\right\}$. The received signals at UEs $i$ and $j$ are given by, 
\vspace{-.2cm}
\begin{alignat}{1}
y_{i}  & =h_{i}\left(\sqrt{p_{i}}x_{A}+\sqrt{p_{j}}x_{B}\right)+z_{i},\\
y_{j}  & =h_{j}\left(\sqrt{p_{i}}x_{A}+\sqrt{p_{j}}x_{B}\right)+z_{j}.\nonumber 
\end{alignat}
For Baseline 2, the same capacity region $\mathcal{R}_{\mathrm{NOMA}}(\mathcal{P}_{\mathrm{NOMA}}) \!=\!\! \bigcup_{\mathbf{p}\in\mathcal{P}_{\mathrm{NOMA}}}\!\! \left\{ (r_{i},r_{j}) \left| \!\! \begin{array}{c} 
r_{i}\le C\big(\frac{p_{i}}{\alpha_{i}}\big), r_{j}\le C\big(\frac{p_{j}}{p_{i}+\alpha_{j}}\big)\end{array} \!\!\!\right.\right\} \label{eq:rate-I}
$ is achieved by SIC with and without caching \cite{Tse2005Fundamentals}, where $x_{B}$ is decoded and canceled before decoding $x_{A}$ at UE $i$. %
{{For Baselines 1 and 2, the rate and time/power allocation is optimized for minimization of the delivery time in a similar manner as for the proposed cache-aided NOMA scheme.}}

\subsection{Simulation Results}

\begin{figure}[t]
\centering\includegraphics[width=3.0in]{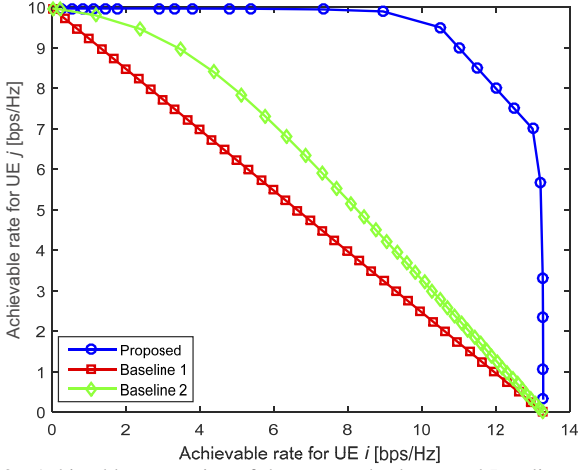}
\vspace{-.4cm}
\caption{\label{fig3} Achievable rate region of the proposed scheme and Baselines 1 and 2 for $\alpha_i = 10^{-3}$ and $\alpha_j = 10^{-2}$.}
\end{figure}

In Fig.~\ref{fig3}, we compare the achievable rate regions of the proposed cache-aided NOMA scheme and the baseline schemes for $\alpha_i = 10^{-3}$ and $\alpha_j = 10^{-2}$. For the proposed scheme, the rate achievable by UE $k$ is given by $r_{k,1} + r_{k,2}$, $k \in \{i, j\}$. Note that the achievable rate regions of all considered schemes are independent of the values of $c_{kf}$, $(k,f)\in \{(i,A), (j,B)\}$. In particular, Baseline 2 with and without caching achieves the same rate region. From Fig.~\ref{fig3}, we observe that all considered schemes achieve the same corner points $(0, 10.0)$ and $(13.2,0)$, since the maximal rate for each UE is fundamentally limited by its channel status. Baseline 1 achieves the smallest rate region as it employs OMA to avoid interference. As NOMA introduces additional degrees of freedom for the users, Baseline 2 has a larger achievable rate region than Baseline 1. The expansion of the rate region is more significant for the weak user than for the strong user since the strong user consumes a small transmit power, and hence, causes little interference to the weak user. The proposed cache-aided NOMA scheme achieves the largest rate region among all considered schemes as joint CIC and SIC allows more interference to be canceled compared to Baseline 2 which can only perform SIC. This translates into a large sum rate gain for the proposed scheme. With the proposed scheme, significant performance gains are possible for both the weak and the strong user due to CIC. 

\begin{figure}[t]
\centering\includegraphics[width=3.0in]{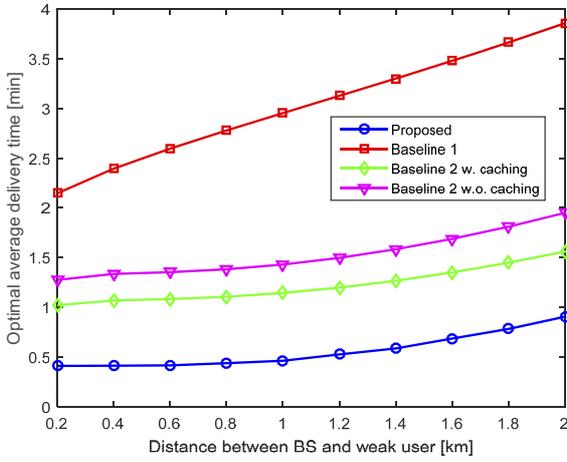}
\vspace{-.4cm}
\caption{\label{fig8} Optimal average delivery time versus the distance between the weak user and the BS.}
\end{figure}

In Fig.~\ref{fig8}, we show the optimal average delivery times of the proposed cache-aided NOMA scheme and Baselines 1 and 2 as functions of the distance of the weak UE to the BS $R_j$. The performance is averaged over different realizations of the user locations and the channel fading. For a given $R_j$, as expected from the achievable rate region results in Fig.~\ref{fig3}, Baseline 1 requires the longest time to complete video file delivery. The proposed cache-aided NOMA scheme outperforms both Baseline 2 without caching and Baseline 2 with caching. This is due to the exploitation of  CIC, which is possible only with the proposed joint caching and NOMA transmission design. However, different from the achievable rate region, the delivery times of Baseline 1, Baseline 2 with caching, and the proposed scheme critically depend on the amount of cached data. As $R_j$ increases, UE $j$, the weak user, suffers from an increased path loss, which in turn reduces the channel gain of UE $j$. Moreover, since the delivery time of the weak user dominates the overall delivery time, we observe from Fig.~\ref{fig8} that the optimal delivery time increases with $R_j$ for all considered schemes. However, Baseline 1 is the least efficient among the considered schemes, and its delivery time increases by about $80$\% as $R_j$ increases from $0.2$~km to $2$~km. By exploiting NOMA and the resulting increased degrees of freedom, Baseline 2 effectively reduces the performance degradation caused by the weak user. {{For example, even without caching, the delivery time of Baseline 2 is $40$\% ($50$\%) lower than that of Baseline 1 when UE $j$ is located at $R_j = 0.2$~km ($R_j = 2$~km). Moreover, when a cache is available, Baseline 2 can also exploit caching for offloading of the delivery data, which further reduces the delivery time compared to Baseline 1 by an additional $12$\% ($10$\%) for $R_j = 0.2$~km ($R_j = 2$~km). The proposed scheme enjoys the best performance and its delivery time is about $80$\% lower than that of Baseline 1 for the considered values of $R_j$.}}

\section{\label{sec6}Conclusion}
In this paper, a joint caching and NOMA transmission design was presented for spectrally efficient downlink communication. The proposed scheme exploits unrequested cached data for cancellation of NOMA interference, which is not possible with separate caching and NOMA transmission. The achievable rate region of the proposed cache-aided NOMA scheme was characterized, and the optimal decoding order and the optimal power and rate allocations for minimization of the delivery time were investigated. Simulation results showed that the proposed scheme can significantly expand the achievable downlink rate region for both the strong and the weak users. Moreover, the delivery time of both users can be effectively reduced to achieve fast video delivery. For ease of illustration, the proposed cache-aided NOMA was only evaluated for the important case of two paired NOMA users. The extension of cache-aided NOMA to multiple users will be considered in future work.

\vspace{.5cm}
\appendices{}
\section{\label{proof1}Proof of Proposition \ref{prop1}}

As $i\overset{(1)}{\to}x_{A1}$, $r_{i,1}\le C_{i,1}$ is achievable for decoding $x_{A1}$ at UE $i$. To derive the achievable rate region, we need to check the decodability of the interfering signals $x_{B2}$ and $x_{A2}$ at UE $i$ and $j$, respectively. Let us consider the following two power regions. 

(1) For $\mathbf{p}\in \mathcal{P}_{1} \backslash\mathcal{P}_{2} $, we have $C\left(\frac{p_{i,2}}{p_{j,1}+\alpha_{j}}\right)<C\left(\frac{p_{i,2}}{p_{j,2}+\alpha_{i}}\right)$, i.e., UE $j$ cannot decode $x_{A2}$ before decoding $x_{B1}$ as the SIC decoding condition is not met. Also, for any $(p_{j,1},p_{j,2})\in\mathbb{R}_{+}^{2}$, $x_{A2}$ cannot be decoded before decoding $x_{B2}$ at UE $j$ as
\vspace{-.2cm}
\begin{alignat}{1}
C \! \left(\! \frac{p_{i,2}}{p_{j,1}+p_{j,2}+\alpha_{j}} \! \right) \!<\! C \! \left(\! \frac{p_{i,2}}{p_{j,2}+\alpha_{j}}\! \right) \!<\! C \! \left(\! \frac{p_{i,2}}{p_{j,2}+\alpha_{i}} \!\right) \!\! .
\label{eq:26-1}
\end{alignat}
On the other hand, for any $(p_{j,1},p_{j,2})\in\mathbb{R}_{+}^{2}$, $x_{B2}$ can be always decoded and canceled at UE $i$ before $x_{A2}$ is decoded as $\alpha_{i}<\alpha_{j}$; and hence, $r_{i,2}\le C_{i,2}$ is achievable. In contrast,  UE $j$ cannot decode $x_{A2}$ in any case. Consequently, the feasible decoding orders are UE $i\overset{(2)}{\to}x_{B2}\overset{(3)}{\to}x_{A2}$ and UE $j\overset{(1)}{\to}(x_{B1}, x_{B2})$, whereby rate region $\mathcal{R}_{1}\left(\mathcal{P}_{1} \backslash\mathcal{P}_{2} \right)$ is achieved.

(2) For $\mathbf{p}\in\mathcal{P}_{2}$, we have $C\left(\frac{p_{i,2}}{\alpha_{i}}\right)>C\left(\frac{p_{i,2}}{p_{j,1}+\alpha_{j}}\right)>C\left(\frac{p_{i,2}}{p_{j,2}+\alpha_{i}}\right)$, i.e., $x_{A2}$ can be decoded at UE $j$ before  $x_{B1}$ is decoded if and only if UE $i \overset{(2)}{\to} x_{A2}$. Assume $x_{A2}$ is decoded last at UE $i$ such that UE $j$ cannot decode $x_{A2}$ in any case. Then, rate region $\mathcal{R}_{1}\left(\mathcal{P}_{2}\right)$ is achievable. On the other hand, suppose $x_{A2}$ is decoded first at UE $i$. Then, UE $j$ can achieve a higher rate for $r_{j,1}$ by decoding $x_{A2}$ before decoding $x_{B1}$, which is only possible after $x_{B2}$ has been decoded according to \eqref{eq:26-1}. Thus, the rate region $\mathcal{R}_{2}\left(\mathcal{P}_{2}\right)$ is achievable.

Therefore, the rate region $\mathcal{R}_{1}\left(\mathcal{P}_{1}\right)\bigcup\mathcal{R}_{2}\left(\mathcal{P}_{2}\right)$ is achievable, and any rate vector outside the region $\mathcal{R}_{1}\left(\mathcal{P}_{1}\right)\bigcup\mathcal{R}_{2}\left(\mathcal{P}_{2}\right)$ cannot be achieved by SIC decoding. This completes the proof.

\vspace{-0.1cm}
\section{\label{proof2} Proof of Proposition \ref{prop2}}
By decoding $x_{B1}$ first, $r_{j,1}\le C\left(\frac{p_{j,1}}{p_{i,2}+p_{j,2}+\alpha_{j}}\right)$ is achievable for UE $j$. To obtain the achievable rate region, two power regions have to be considered. 

(1) For $\mathbf{p}\in\mathcal{P}_{3}$, we have 
\vspace{-.2cm}
\begin{alignat}{1}
C\left(\frac{p_{i,2}}{p_{i,1}+p_{j,2}+\alpha_{i}}\right)>C\left(\frac{p_{i,2}}{p_{j,2}+\alpha_{j}}\right),\label{eq:III-1}\\
C\left(\frac{p_{j,2}}{p_{i,1}+p_{i,2}+\alpha_{i}}\right)>C\left(\frac{p_{j,2}}{p_{i,2}+\alpha_{j}}\right),\label{eq:III-2}
\end{alignat}
which imply that $x_{A2}$ cannot be decoded at UE $j$ in general, cf.~\eqref{eq:III-1}, but $x_{B2}$ can always be decoded at UE $i$, cf. \eqref{eq:III-2}. Consequently, UE $j$ can decode $x_{B2}$ only by treating $x_{A2}$ as noise whereas UE $i$ will first decode $x_{B2}$ and cancel its contribution to the received signal before decoding $x_{A1}$ and $x_{A2}$. Therefore, the achievable rate region is given by $\mathcal{R}_{3}\left(\mathcal{P}_{3}\right)$.

(2) For $\mathbf{p}\in \mathcal{P}_{4}$, we have
\vspace{-.2cm}
\begin{alignat}{1}
C\left(\frac{p_{i,2}}{p_{i,1}+p_{j,2}+\alpha_{i}}\right) & <C\left(\frac{p_{i,2}}{p_{j,2}+\alpha_{j}}\right),\label{eq:III-3}\\
C\left(\frac{p_{j,2}}{p_{i,1}+p_{i,2}+\alpha_{i}}\right) & <C\left(\frac{p_{j,2}}{p_{i,2}+\alpha_{j}}\right),\label{eq:III-4}\\
C\left(\frac{p_{j,2}}{p_{i,1}+\alpha_{i}}\right) & <C\left(\frac{p_{j,2}}{\alpha_{j}}\right).\label{eq:III-5}
\end{alignat}
That is, at UE $i$, $x_{B2}$ cannot be decoded first, cf. \eqref{eq:III-4}. Hence, we only need to consider UE $i\overset{(1)}{\to}x_{A2}$. In this case, UE $j$ is able to cancel the interference from $x_{A2}$ before decoding $x_{B2}$ due to \eqref{eq:III-3}. However, at UE $i$, $x_{B2}$ cannot be canceled before decoding $x_{A1}$ due to \eqref{eq:III-5}, i.e., UE $i\overset{(2)}{\to}x_{B2}$ is infeasible. Therefore, the achievable rate region is given by $\mathcal{R}_{4}\left(\mathcal{P}_{4}\right)$, where UE $i$ cannot decode $x_{B2}$ while UE $j$ can decode and cancel $x_{A1}$ before decoding $x_{B2}$. 
%
%
%
%
Therefore, the rate region in Proposition~\ref{prop2} 
is achievable, which completes the proof.

\vspace{-0.1cm}
\section{\label{proof3}Proof of Proposition \ref{prop3}}

First, assume UE $j\overset{(1)}{\to}x_{B2}$ and UE $i\overset{(1)}{\to}(x_{A2},x_{B2})$. If $\mathbf{p}\in\mathcal{P}_{5}$, we have
\vspace{-.2cm}
\begin{alignat}{1}
 & C\left(\frac{p_{j,2}}{p_{i,1}+p_{i,2}+\alpha_{i}}\right)>C\left(\frac{p_{j,2}}{p_{i,2}+p_{j,1}+\alpha_{j}}\right),\label{eq:32}\\
 & C\left(\frac{p_{i,2}}{\alpha_{i}}\right)>C\left(\frac{p_{i,2}}{p_{i,1}+\alpha_{i}}\right)>C\left(\frac{p_{i,2}}{p_{j,1}+\alpha_{j}}\right).\label{eq:33}
\end{alignat}
By \eqref{eq:32}, UE $i$ can decode and cancel $x_{B2}$ as $\alpha_{i}<\alpha_{j}$. Thus, UE $i\overset{(1)}{\to}x_{B2}$, which leads to the residual received signals $y_{i}^{(1)}=h_{i}\left(\sqrt{p_{i,1}}x_{A1}+\sqrt{p_{i,2}}x_{A2}\right)+z_{i}$ and $y_{j}^{\mathrm{(1)}}=h_{j}\left(\sqrt{p_{j,1}}x_{B1}+\sqrt{p_{i,2}}x_{A2}\right)+z_{j}$. By \eqref{eq:33}, UE $j$ cannot decode $x_{A2}$ based on $y_{j}^{(1)}$. Therefore, the decoding orders UE $i\overset{(1)}{\to}x_{B2}\overset{(2)}{\to}(x_{A1},x_{A2})$ and UE $j\overset{(1)}{\to}x_{B2}\overset{(2)}{\to}x_{B1}$ are feasible and  achieve rate region $\mathcal{R}_{5}\left(\mathcal{P}_{5}\right)$.

However, if $\mathbf{p}\in \mathcal{P}_{6}$, UE $i$ cannot decode $x_{B2}$ first due to \eqref{eq:32}. Then, for the assumption of UE $i\overset{(1)}{\to}(x_{A2},x_{B2})$, we only need to consider the case UE $i\overset{(1)}{\to}x_{A2}$. 
We have 
\vspace{-.2cm}
\begin{alignat}{1}
C\left(\frac{p_{i,2}}{p_{j,1}+\alpha_{j}}\right)>C\left(\frac{p_{i,2}}{p_{i,1}+\alpha_{i}}\right),\label{eq:38}
\end{alignat}
i.e., UE $j$ can cancel $x_{A2}$ before decoding $x_{B1}$. On the other hand, UE $i$ cannot cancel $x_{B2}$ before decoding $x_{A1}$ unless $\Delta=0$, whereby we have $C\left(\frac{p_{j,2}}{p_{i,2}+p_{j,1}+\alpha_{j}}\right) < C\left(\frac{p_{j,2}}{p_{i,1}+\alpha_{i}}\right)$. Hence, rate region $\mathcal{R}_{6}\left(\mathcal{P}_{6}\right)$ is achievable.

Next, assume UE $j\overset{(1)}{\to}x_{A2}$ and 
UE $i\overset{(1)}{\to}x_{A2}$, which requires $C\left(\frac{p_{i,2}}{p_{j,1}+p_{j,2}+\alpha_{j}}\right)>C\left(\frac{p_{i,2}}{p_{i,1}+p_{j,2}+\alpha_{i}}\right)$, or equivalently, $\mathbf{p}\in \mathcal{P}_{7}$. In this case, as $C\left(\frac{p_{j,2}}{\alpha_{i}}\right)>C\left(\frac{p_{j,2}}{p_{j,1}+\alpha_{j}}\right)>C\left(\frac{p_{j,2}}{p_{i,1}+\alpha_{i}}\right)$, UE $i$ cannot decode $x_{B2}.$ Hence, the decoding orders UE $i\overset{(1)}{\to}x_{A2}\overset{(2)}{\to}x_{A1}$ and UE $j \overset{(1)}{\to} x_{A2}\overset{(2)}{\to}(x_{B1},x_{B2})$ are feasible and achieve rate region $\mathcal{R}_{7}\left(\mathcal{P}_{7}\right)$.

Finally, for UE $j\overset{(1)}{\to}x_{A2}$ and UE $i\overset{(1)}{\to}x_{B2}$, feasible power and rate allocations do not exist. In particular, for such a rate region to exist, the following inequalities would have to hold,
\vspace{-.2cm}
\begin{alignat}{1}
C\left(\frac{p_{i,2}}{p_{j,1}+p_{j,2}+\alpha_{j}}\right) & >C\left(\frac{p_{i,2}}{p_{i,1}+\alpha_{i}}\right),\label{eq43-1}\\
C\left(\frac{p_{j,2}}{p_{i,1}+p_{i,2}+\alpha_{i}}\right) & >C\left(\frac{p_{j,2}}{p_{j,1}+\alpha_{j}}\right),\label{eq43-2}
\end{alignat}
which ensure feasibility of UE $j\overset{(1)}{\to}x_{A2}$ and UE $i\overset{(1)}{\to}x_{B2}$, respectively. Eqs. \eqref{eq43-1} and \eqref{eq43-2} are equivalent to $p_{i,1}-p_{j,1}>p_{j,2}+\alpha_{j}-\alpha_{i}$ and $p_{i,1}-p_{j,1}<\alpha_{j}-\alpha_{i}-p_{j,2}$, respectively, which lead to $p_{i,2}+p_{j,2}<0$. That is, \eqref{eq43-1} and \eqref{eq43-2} cannot be met for feasible powers. Therefore, the rate region in Proposition~\ref{prop3} 
is achievable, which completes the proof.

\end{document}